\documentclass[epj]{svjour}
\usepackage{latexsym}
\usepackage{graphics}
\begin{document}
\title{Flavor Structure of the Nucleon Sea}
\author{Jen-Chieh Peng
}                     
\institute{University of Illinois at Urbana-Champaign, Urbana 61801 IL USA}
\date{Received: date / Revised version: date}
%
\abstract{
The recent progress on our understanding of the flavor structure of
unpolarized and polarized nucleon sea is reviewed. The large flavor 
asymmetry between the up and down sea quark distributions is now well
established. This asymmetry strongly suggests the importance of the
mesonic degrees of freedom in the description of the nucleon sea.
The strong connection between the flavor structure and the spin structure
of the nucleon sea is emphasized. Possible future measurements 
for testing various theoretical models are also discussed.
\PACS{
      {14.20.Dh}{Properties of protons and neutrons} \and
      {24.85.+p}{Quarks, gluons, and QCD in nuclei and nuclear processes}
     } 
} 
\maketitle
\section{Introduction}
\label{intro}

One of the most active areas of research in nuclear and particle physics 
during the last several decades is the study of quark and gluon distributions 
in the nucleons and nuclei. Several major surprises were discovered in 
Deep-Inelastic Scattering (DIS) experiments which profoundly changed our 
views of the partonic substructure of hadrons. In the early 1980's, the 
famous `EMC' effect found in muon DIS provided the first 
unambiguous evidence that the quark distributions in nuclei are
significantly different from those in free nucleons. 
More recently, surprising results on the spin and
flavor structures of the nucleons were discovered in DIS experiments. 
The so-called ``spin crisis'', revealed by the polarized DIS 
experiments, has led to extensive efforts to 
understand the partonic content of proton's 
spin. Subsequently, the observation~\cite{nmc91} of the 
violation of the Gottfried sum rule~\cite{gott} in DIS revealed a 
surprisingly large asymmetry between the up and down antiquark distributions 
in the nucleon, shedding new light on the origins of the nucleon sea.

In this article, we review the status of our current knowledge on the
flavor dependence of the sea quark distributions in hadrons. 
In Section 2, we summarize the 
experimental evidence for the flavor asymmetry of the nucleon sea.
Implications of various theoretical models for 
explaining the $\bar d/ \bar u$ asymmetry are also discussed.
Section 3 is devoted to the subject of the flavor structure of polarized
nucleon sea. Finally, we present future prospects and conclusion in
Section 4.

\section{Flavor structure of unpolarized nucleon sea}
\label{sec:1}

The earliest parton models assumed that the proton sea was flavor symmetric,
even though the valence quark distributions are clearly flavor asymmetric.
Inherent in this assumption is that the content of the sea is 
independent of the valence quark's composition.
The flavor symmetry assumption was not based on any known physics, and 
it remained to be tested by experiments. Neutrino-induced charm production
experiments~\cite{abrom82,conrad98}, which are 
sensitive to the $s \to c$ process, showed 
that the strange-quark content of the nucleon is only about half
of the up or down sea quarks. Such flavor asymmetry is attributed to the
much heavier strange-quark mass compared to the up and down quarks. The similar
masses for the up and down quarks suggest that the nucleon sea should be nearly
up-down symmetric. 

The issue of the equality of $\bar u$ and $\bar d$ was first
encountered in measurements of the Gottfried integral~\cite{gott},
defined as
\begin{equation}
I_G = \int_0^1 \left[F^p_2 (x,Q^2) - F^n_2 (x,Q^2)\right]/x~ dx,
\end{equation}
where $F^p_2$ and $F^n_2$ are the proton and neutron structure
functions measured in DIS
experiments.
Under the assumption of a $\bar u$, $\bar d$ flavor-symmetric sea in
the nucleon, the Gottfried Sum Rule (GSR)~\cite{gott}, $I_G
= 1/3$, is obtained.
The most accurate test of the GSR was reported by the New Muon 
Collaboration (NMC)~\cite{nmc91}, which measured $F^p_2$ and $F^n_2$ over the 
region $0.004 \le x \le 0.8$. They determined the Gottfried integral to be 
$ 0.235\pm 0.026$, significantly below 1/3. This surprising result has
generated much interest. Although the violation of the GSR can be 
explained by assuming unusual behavior of the parton distributions at very
small $x$, a more natural explanation is to abandon the assumption
$\bar u = \bar d$.

The proton-induced Drell-Yan (DY) process provides an
independent means to probe the flavor asymmetry of the nucleon sea~\cite{es}.
An important advantage of the DY process is that the $x$ dependence of 
$\bar d / \bar u$ can be determined.
The NA51 collaboration at CERN carried out the first dedicated dimuon
production experiment to study the flavor structure of the 
nucleon sea~\cite{na51}. Using a 450 GeV proton beam, NA51 obtained 
$\bar u/\bar d = 0.51 \pm 0.04 (stat) \pm 0.05 (syst)$
at $x = 0.18$ and $\langle M_{\mu\mu}\rangle = 5.22$ GeV. This 
important result established
the asymmetry of the quark sea at a single value of $x$. What remained
to be done was to map out the $x$-dependence of this asymmetry. 

\begin{figure}
\resizebox{0.45\textwidth}{!}{%
  \includegraphics{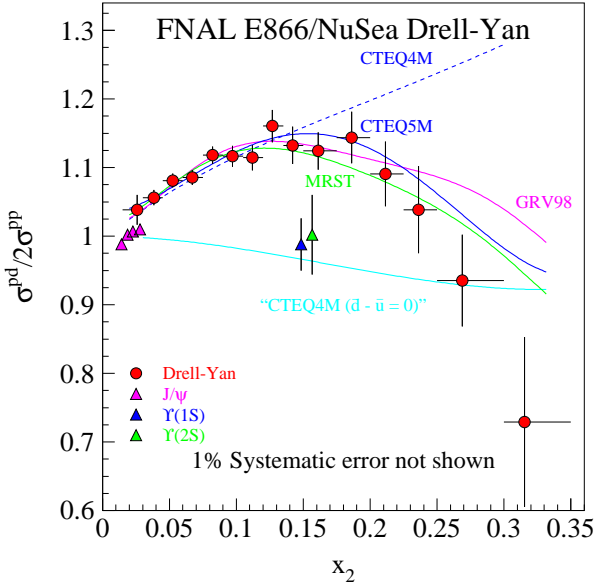}
}
\caption{Cross section ratios of $p+d$ over $2(p+p)$ for Drell-Yan, $J/\Psi$,
and $\Upsilon$ production from FNAL E866. The curves are the calculated 
next-to-leading-order cross section ratios for the Drell-Yan using various
parton distribution functions.}
\label{fig:1}       
\end{figure}

At Fermilab, a DY experiment (E866/NuSea) aimed at a higher statistical
accuracy with a much wider kinematic coverage than the NA51 experiment
has been completed~\cite{e866,peng,towell}. The DY cross section ratio
per nucleon for $p + d$ to that for $p + p$ is 
shown in Fig.~\ref{fig:1}.
At positive $x_F$ this ratio is given as 
\begin{equation}
\sigma_{DY}(p+d)/2\sigma_{DY}(p+p) \simeq
(1+\bar d(x_2)/\bar u(x_2))/2.
\label{eq:3.3}
\end{equation}
Figure~\ref{fig:1} shows that the DY cross section per nucleon for 
$p + d$ clearly exceeds $p + p$, and it indicates an excess of $\bar d$ 
with respect to $\bar u$ over an appreciable range in $x_2$.
In contrast, the $\sigma(p+d)/2\sigma(p+p)$ ratios for $J/\Psi$ and
$\Upsilon$ production, also shown in Fig. 1, are very close to unity.
This reflects the dominance of gluon-gluon fusion process for
quarkonium production and the expectation that the gluon distributions
in the proton and in the neutron are identical. 

The Drell-Yan cross section ratios from E866 were analysed to obtain
$\bar d - \bar u$ over the region $0.02 < x < 0.345$ as shown in Fig. 2.
The HERMES collaboration has reported a semi-inclusive
DIS measurement of charged pions from hydrogen and deuterium 
targets~\cite{hermes}.
Based on the differences between charged-pion yields from the two targets,
$\bar d - \bar u$ is determined
in the kinematic range, $0.02 < x < 0.3$ and 
1 GeV$^2$/c$^2 < Q^2 <$ 10 GeV$^2$/c$^2$. The HERMES results 
are consistent with
the E866 results obtained at significantly higher $Q^2$.
In Table 1 we list the
values of the integral $\int_0^1 [\bar d(x) - \bar u(x)] dx$ determined
from the NMC, HERMES, and FNAL E866 experiments. The agreement among these
results, obtained using different techniques including DIS, semi-inclusive
DIS, and Drell-Yan, is quite good.
\begin{figure}
\resizebox{0.45\textwidth}{!}{%
  \includegraphics{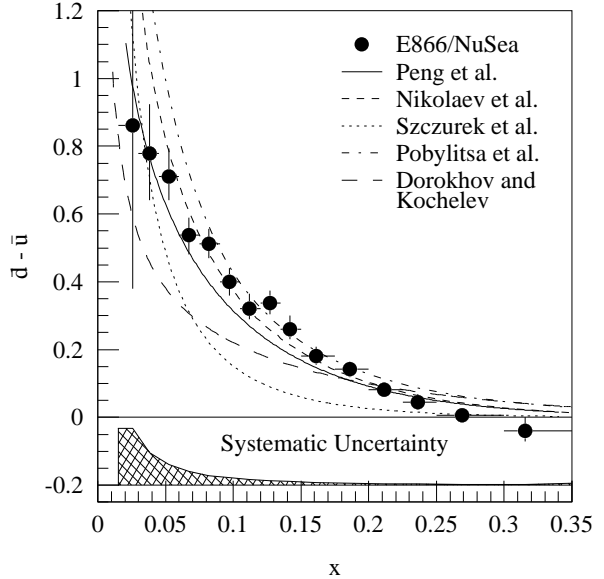}
}
\caption{Comparison of the measured $\bar d(x) - \bar u(x)$ at $Q^2 = 54$
GeV$^2$/c$^2$ to predictions of several models of the nucleon sea.
The solid and short-dashed curves show pion-cloud 
calculations~\cite{peng,nikolaev}. The dotted
curve is a chiral quark model calculation~\cite{szczurek}, while 
the dot-dash curve shows the
chiral quark-soliton calculation~\cite{pobylitsa}. 
The long-dash curve shows the instanton 
model prediction~\cite{dorokhov}.}
\label{fig:2}       
\end{figure}
\begin{table}
\caption{Values of the integral 
$\int_0^1 [\bar d(x) - \bar u(x)] dx$ determined from the DIS, 
semi-inclusive DIS, and Drell-Yan experiments.}
\label{tab:1}       
\begin{tabular}{lll}
\hline\noalign{\smallskip}
Experiment & $\langle Q^2 \rangle$ (GeV$^2$/c$^2$) & 
$\int_0^1 [\bar d(x) - \bar u(x)] dx$  \\
\noalign{\smallskip}\hline\noalign{\smallskip}
NMC/DIS & 4.0 & $0.147 \pm 0.039$ \\
HERMES/SIDIS & 2.3 & $0.16 \pm 0.03$ \\
FNAL E866/DY & 54.0 & $0.118 \pm 0.012$ \\
\noalign{\smallskip}\hline
\end{tabular}
\end{table}

Many theoretical models, including meson cloud model, chiral-quark 
model, Pauli-blocking model, instanton model, chiral-quark soliton model,
and statistical model, have been proposed to explain the $\bar d/ \bar u$
asymmetry. For details of these various models, we refer to several
recent review articles~\cite{kumano98,speth98,garvey02}. 
As shown in Fig.~\ref{fig:2}, these models can describe 
the $\bar d - \bar u$ data very well. However, they all have 
difficulties explaining the $\bar d / \bar u$ data at 
large $x$ ($x>0.2$)~\cite{melnitchouk}.
Thus, it would be very important to extend the DY
measurements to larger $x_2$ regimes.
The new 120 GeV Fermilab Main Injector (FMI)
and the proposed 50 GeV 
Japanese Hadron Facility~\cite{nagamiya} (JHF) present opportunities for 
extending the $\bar d/ \bar u$ measurement to larger $x$ ($0.25 < x < 0.7$).
Figure 3 shows the expected statistical 
accuracy for $\sigma (p+d)/ 2 \sigma (p+p)$ at 
JHF~\cite{peng00} compared 
with the data from E866 and a proposed measurement~\cite{e906} using 
the 120 GeV proton beam at the FMI. A definitive measurement of
the $\bar d/ \bar u$ over the region $0.25 < x < 0.7$ could indeed be
obtained at FMI and JHF.

\begin{figure}
\resizebox{0.45\textwidth}{!}{%
  \includegraphics{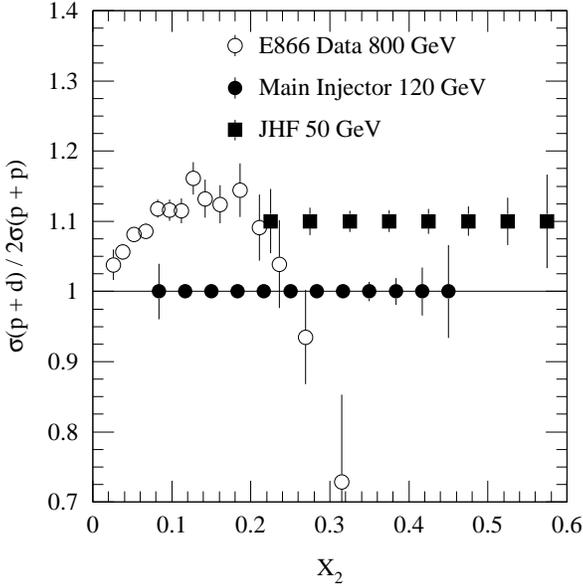}
}
\caption{
Projected statistical accuracy for $\sigma (p+d) /2\sigma (p+p)$ in a
100-day run at JHF~\cite{peng00}. The E866 
data and the projected sensitivity for
a proposed measurement~\cite{e906} at the 120 GeV
Fermilab Main-Injector are also shown.}
\label{fig:3}       
\end{figure}

To disentangle the $\bar d / \bar u$ asymmetry from the possible 
charge-symmetry violation effect~\cite{ma1,boros}, 
one could consider $W$ boson production in $p + p$ collision at RHIC.
An interesting quantity to be measured is the ratio of the 
$p + p \to W^+ + x$ and $p + p \to W^- + x$ cross sections~\cite{peng1}. 
It can be shown that this
ratio is very sensitive to $\bar d / \bar u$. An important feature of
the $W$ production asymmetry in $p + p$ collision 
is that it is completely free 
from the assumption of charge symmetry. Figure 4 shows the 
predictions for $p + p$ collision at $\sqrt s =
500~$GeV. The dashed curve corresponds to the $\bar
d/\bar u$ symmetric MRS S0$^\prime$ structure 
functions, while the solid and dotted curves
are for the $\bar d/\bar u$ asymmetric structure function MRST and MRS(R2),
respectively. Figure 4 clearly shows that $W$ asymmetry 
measurements at RHIC could provide an independent determination 
of $\bar d / \bar u$.
\begin{figure}
\resizebox{0.45\textwidth}{!}{%
  \includegraphics{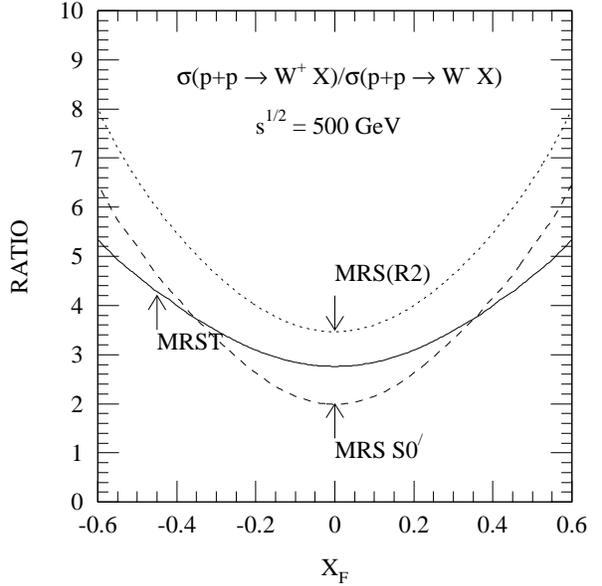}
}
\caption{Predictions of
$\sigma (p+p \to W^+ x) / \sigma (p+p \to W^- x)$ as a function of $x_F$
at $\sqrt s$ = 500 GeV. The dashed curve corresponds to the $\bar
d/\bar u$ symmetric MRS S0$^\prime$ structure
functions, while the solid and dotted curves
are for the $\bar d/\bar u$ asymmetric structure function MRST and MRS(R2),
respectively.}
\label{fig:4}       
\end{figure}

Models in which virtual mesons are admitted as degrees of freedom
have implications that extend beyond the $\bar d, \bar u$ 
flavor asymmetry addressed above.
They create hidden strangeness in the nucleon via such virtual processes as 
$p \to \Lambda + K^+, \Sigma + K$, etc. 
Such processes are of considerable interest as they imply different $s$ and
$\bar s$ parton distributions in the nucleon, a feature not found in gluonic
production of $s \bar s$ pairs. This subject has 
received a fair amount of attention
in the literature~\cite{signal87,ji95,ma96} 
but experiments have yet to clearly identify such a
difference. Thus in contrast to the $\bar d, \bar u$
flavor asymmetry, to date there is no positive experimental evidence
for $s \bar s$ contributions to the nucleon from virtual 
meson-baryon states~\cite{ccfr95,barone99,goncharov01}.

A difference between the $s$ and $\bar s$ distribution can be made manifest
by direct measurement of the $s$ and $\bar s$ parton distribution functions 
in DIS neutrino scattering, or in
the measurement of the $q^2$ dependence of the 
strange quark contribution ($F^p_{1s}(q^2)$)
to the proton charge form factor. 
Measurement of these form
factors allows extraction of the strangeness contribution to the 
nucleon's charge and magnetic moment and axial form factors. The portion 
of the charge form factor $F^p_{1s} (q^2)$ due to strangeness clearly is
zero at $q^2 = 0$, but if the $s$ and $\bar s$ distributions are different the
form factor becomes non-zero at finite $q^2$. These ``strange'' form 
factors can be measured in neutrino elastic scattering~\cite{garvey93}
from the nucleon, or by selecting the parity-violating component of
electron-nucleon elastic scattering, as is now
being done at the Bates~\cite{mueller} and Jefferson Laboratories~\cite{aniol}.

\section{Flavor structure of polarized nucleon sea}
\label{sec:3}

The flavor structure and the 
spin structure of the nucleon sea are closely connected. Many 
theoretical models originally proposed to explain 
the $\bar d / \bar u$ flavor asymmetry also have specific implications for
the spin structure of the nucleon sea. In the meson-cloud model, for example,
a quark would undergo a spin flip upon an
emission of a pseudoscalar meson ($
u^\uparrow \to \pi^\circ (u \bar u, d \bar d) + u^\downarrow,~u^\uparrow
\to \pi^+ (u \bar d) + d^\downarrow,~u^\uparrow \to K^+ + s^\downarrow$,
etc.). The antiquarks ($\bar u, \bar d, \bar s$) are 
unpolarized ($\Delta \bar u = \Delta \bar d = \Delta \bar s = 0$)
since they reside in spin-0
mesons. The strange quarks ($s$), on the other hand, would have a negative 
polarization since the up valence quarks in the proton 
are positively polarized and
the $u^\uparrow \to K^+ + s^\downarrow$ process would lead to an excess
of $s^\downarrow$. By considering a vector meson ($\rho$) cloud, 
non-zero $\bar u, \bar d$ sea quark polarizations with 
$\Delta \bar d - \Delta \bar u > 0$ were 
predicted~\cite{fries98,boreskov99,cao01,kumano01}.

The Pauli-blocking model~\cite{steffens} predicts that an excess of 
$q^\uparrow (q^\downarrow)$ valence quarks
would inhibit the creation of a pair of $q^\uparrow \bar q^\downarrow$
($q^\downarrow \bar q^\uparrow$) sea quarks. Since the polarization
of the $u$($d$) valence quarks are positive (negative), this
model predicts a positive (negative) polarization for the $\bar u
(\bar d)$ sea $(\Delta \bar u > 0 > \Delta \bar d)$.

In the instanton model~\cite{dorokhov}, the 
quark sea can result from a scattering of a 
valence quark off a nonperturbative vacuum fluctuation of the gluon field,
instanton. The interaction induced by an instanton is given by the 't Hooft
effective lagrangian which allows processes such as $u^\uparrow \to 
u^\downarrow d^\uparrow \bar d^\downarrow$, $d^\downarrow \to d^\uparrow 
u^\downarrow \bar u^\uparrow$, etc. Since the flavor of the quark-antiquark 
produced in this process is different from the flavor of the initial valence
quark, this model readily explains $\bar d > \bar u$. Furthermore,
the correlation between the sea quark helicity and the valence quark
helicity in the 't Hooft effective lagrangian (i.e. $u^\uparrow$ leads to a
$\bar d^\downarrow$) naturally predicts a positively (negatively) polarized
$\bar u (\bar d)$ sea. In particular, this model predicts~\cite{dorokhov01} 
a large $\Delta \bar u, \Delta \bar d$ flavor asymmetry with 
$\Delta \bar u > \Delta \bar d$, namely,
$\int_0^1 [\Delta \bar u(x) - \Delta \bar d(x)] dx = \frac{5}{3} \int_0^1
[\bar d(x) - \bar u(x)] dx.$

In the chiral-quark soliton model~\cite{diakonov96,wakamatsu98}, 
the large $N_c$ limit of 
QCD becomes an effective theory of mesons with the baryons 
appearing as solitons. Quarks are described by single particle wave 
functions which are solutions of the Dirac equation in the field of
the background pions. In this model, the polarized isovector distributions
$\Delta \bar u(x) - \Delta \bar d(x)$ appears in leading-order ($N_c^2$)
in a $1/N_c$ expansion, while the unpolarized isovector distributions 
$\bar u(x) - \bar d(x)$ appear in next-to-leading order ($N_c$).
Therefore, this model predicts a large flavor asymmetry for the polarized sea
$[\Delta \bar u (x) - \Delta \bar d(x)] > [\bar d(x) - \bar u(x)]$.

In the statistical approach, the nucleon is treated as a collection of
massless quarks, antiquarks, and gluons in thermal equilibrium within
a finite size volume. The momentum distributions for quarks and
antiquarks follow a Fermi-Dirac distributions function characterized
by a common temperature and a chemical potential $\mu$ which depends on
the flavor and helicity of the quarks. It can be shown that
\begin{equation}
\mu_{\bar q\uparrow} = - \mu_{q\downarrow};~\mu_{\bar q\downarrow} =
- \mu_{q\uparrow}.
\end{equation}
\noindent Eq. 3, together with the constraints of the valence quark
sum rules and inputs from polarized DIS experiments, can readily lead 
to the prediction that $\bar d > \bar u$ and 
$\Delta \bar u > 0 > \Delta \bar d$.
\begin{table}
\caption{Prediction of various theoretical models on the integral
$I_\Delta = \int_0^1 [\Delta \bar u(x) - \Delta \bar d(x)] dx$.}
\label{tab:2}       
\begin{tabular}{lll}
\hline\noalign{\smallskip}
Model & $I_\Delta$ prediction & Ref.  \\
\noalign{\smallskip}\hline\noalign{\smallskip}
Meson cloud & 0 & \cite{ehq,thomas} \\
($\pi$-meson) & & \\
Meson cloud & $\simeq -0.0007$ to $-0.027$ & \cite{fries98} \\
($\rho$-meson) & & \\
Meson cloud & $= -6 \int_0^1 g^p(x) dx$ & \cite{boreskov99} \\
($\pi-\rho$ interf.) & $\simeq -0.7$ & \\
Meson cloud & $\simeq -0.004$ to $-0.033$ & \cite{cao01} \\
($\rho$ and $\pi-\rho$ interf.) & & \\
Meson cloud & $< 0$ & \cite{kumano01} \\
($\rho$-meson) & & \\
Meson cloud & $\simeq 0.12$ & \cite{fries02} \\
($\pi-\sigma$ interf.) & & \\
Pauli-blocking & $\simeq 0.09$ & \cite{cao01} \\
(bag-model) & & \\
Pauli-blocking & $\simeq 0.3$ & \cite{gluck00} \\
(ansatz) & & \\
Pauli-blocking & $= \frac{5}{3} \int_0^1 [\bar d(x) - \bar u(x)] dx$ & \cite{steffens02} \\
  & $\simeq 0.2$ & \\
Chiral-quark soliton & 0.31 & \cite{dressler98} \\
Chiral-quark soliton & $\simeq \int_0^1 2x^{0.12} [\bar d(x) - \bar u(x)] dx$ & \cite{wakamatsu99} \\
Instanton & $= \frac{5}{3} \int_0^1 [\bar d(x) - \bar u(x)] dx$ & \cite{dorokhov01} \\
 & $\simeq 0.2$ & \\
Statistical & $\simeq \int_0^1 [\bar d(x) - \bar u(x)] dx$ & \cite{bourrely01} \\
 & $\simeq 0.12$ & \\
Statistical & $> \int_0^1 [\bar d(x) - \bar u(x)] dx$ & \cite{bhalerao00} \\
 & $> 0.12$ & \\
\noalign{\smallskip}\hline
\end{tabular}
\end{table}

Predictions of various model calculations for $I_\Delta$,
the first moment of
$\Delta \bar u(x) - \Delta \bar d(x)$, are listed in Table 2. 
While the meson cloud model gives small negative values for
$I_\Delta$, all other models predict a positive $I_\Delta$ with
a magnitude comparable or greater than the corresponding integral
for unpolarized sea (recall that $\int_0^1 [\bar d(x) - \bar u(x)] dx \simeq
0.12$). Several meson-cloud calculations for the direct contribution of
$\rho$ meson cloud are in good agreement. However, the large $\pi - \rho$
interference effect reported in ~\cite{boreskov99} was not confirmed
in a more recent study~\cite{cao01}. It is worth noting that a recent 
work~\cite{fries02} 
on $\pi - \sigma$ interference predicts a large effect on 
$\Delta \bar d - \Delta \bar u$, and with a sign opposite to other meson cloud
model calculations.

If the flavor asymmetry of the polarized sea is indeed as large as the
predictions of many models shown in Table 2, it would imply that a
significant fraction of the Bjorken sum, $\int _0^1 [g^p_1(x) - g^n_1(x)] dx$,
comes from the flavor asymmetry of polarized nucleon sea.

Measurements of $\Delta \bar u(x)$ and $\Delta \bar d(x)$ are clearly of 
great current interest. The HERMES collaboration 
has reported their 
preliminary results on the extraction of $\Delta \bar u(x)$ and 
$\Delta \bar d(x)$ using polarized semi-inclusive 
DIS data~\cite{jackson02}. A global analysis
of inclusive spin asymmetries for $\pi^+$, $\pi^-$, $K^+$, and $K^-$ has been
carried out for longitudinally polarized hydrogen and deuterium targets.
As a result, $\Delta u(x)$, $\Delta d(x)$, $\Delta \bar u(x)$, 
$\Delta \bar d(x)$, $\Delta s(x) (=\Delta \bar s(x))$ polarized quark
densities were extracted for $0.03 < x$ at $Q^2 = 2.5$ GeV$^2$.
These very interesting preliminary results showed that $\Delta s$ has
a trend of being positive, in disagreement with the predictions of theoretical
models which attributed the violation of Ellis-Jaffe sum rule to a large
negative polarization of the strange sea. Furthermore, the preliminary 
HERMES result does not support the prediction of a large positive $\Delta
\bar u - \Delta \bar d$. Although the statistics are still limited, the
HERMES preliminary result shows that $\Delta \bar u, \Delta \bar d, \Delta 
\bar u - \Delta \bar d$ are all consistent with being zero.

Another promising technique for measuring sea-quark polarization is
$W$-boson production\cite{bunce00} at RHIC.
The longitudinal single-spin asymmetry for $W$ production in polarized
$ p + p \to W^{\pm} + x$ can be written in leading order as

\begin{equation}
A_L^{W^+} \approx - \frac{\Delta \bar d(x)}{\bar d(x)},~~
A_L^{W^-} \approx - \frac{\Delta \bar u(x)}{\bar u(x)}
\label{Eq:wprod2}
\end{equation}

\noindent at suitable kinematic regions. 
Therefore, $A_L$ gives a direct measure of
sea-quark polarization. The RHIC $W$-production and
the HERMES SIDIS measurements are clearly complementary tools for
determining polarized sea quark distributions.

\section{Conclusion}

The flavor asymmetry of the nucleon sea has been clearly established
by recent DIS and Drell-Yan experiments. 
The $x$ dependence of $\bar d / \bar u$ indicates that a
$\bar d, \bar u$ symmetric sea dominates at small ($x < 0.05$) and 
large $x$ ($x > 0.3$). But for $0.1 < x < 0.2$ a large and significant
flavor non-symmetric contribution determines the sea distributions.
The surprisingly large asymmetry
between $\bar u$ and $\bar d$ is unexplained by perturbative
QCD, and it strongly suggests the presence of virtual isovector 
mesons, mostly pions, in the nucleon sea.
Additional clues on the origins of the flavor asymmetry will come from
future studies including:

\begin{itemize}

\item Measurements of $\bar d/ \bar u$ for $x > 0.25$.

\item Mesurements of $\Delta \bar u$ and $\Delta \bar d$ using semi-inclusive 
DIS, and $W$ production in polarized p-p collision.

\item Direct measurement of the meson cloud in DIS experiments tagging
on forward-going nucleons. Interesting first measurements were performed
recently at HERA~\cite{adloff99}.

\item More precise measurements on the $s$ versus $\bar s$ distributions in
the nucleon.

\end{itemize}

\end{document}